\documentstyle[epsfig,12pt]{article}

\begin{document}

\vspace{1cm}
\begin{center}
\baselineskip=16pt
{\Large\bf Symmetries of Taub-NUT Dual Metrics}
\vskip 2cm
{\bf Dumitru B\u{a}leanu\footnote{after 1st March 1999 at Institute for
Space Sciences, P.O.BOX, MG-36, R 76900, Bucharest-M\u{a}gurele,Romania,
e-mail:baleanu@roifa.ifa.ro\\}~\footnote{E-mail address:
baleanu@thsun1.jinr.ru},
{\bf Sorin Codoban\footnote{E-mail address: codoban@thsun1.jinr.ru}}\\
}
\vskip 0.5cm
{\small \em Joint Institute for Nuclear Research}\\
{\small \em Bogoliubov Laboratory of Theoretical Physics},\\
{\small \em 141 980 Dubna, Moscow Region, Russia}\\
\vskip 0.8cm
\end{center}

\vskip 1 cm
\begin{abstract}
  In this paper we study the symmetries of the dual Taub-NUT
me\-trics.  {\it Generic} and {\it non-generic} symmetries of dual
Taub-NUT me\-trics are investigated.
  The existence of the Runge-Lenz type symmetry  is
analyzed for dual Taub-NUT metrics.
 We find that in some cases the symmetries of the dual metrics
 are the same with the symmetries of Taub-NUT metric.
\end{abstract}
 \bigskip
 \newpage
 %%%%%%%%%%%%%%%%%%%%%%%%%
\section{Introduction}
  In a geometrical setting, symmetries are connected with isometries
associated with Killing vectors and, more generally, Killing tensors
on the configurations space of the system. An example is the motion
of a point particle in a space with isometries \cite{1}, which
is a physicist's way of studying the geodesic structure of a manifold.
In \cite{1} such studies were
extended to spinning space-times described by supersymmetric extensions
of the geodesic motion, and in\cite{2} it was shown that this
can give rise to interesting new types of supersymmetry as well.

 The geometric  duality between a metric
$g^{\mu\nu}$ and its non-degenerate  Killing tensor $K^{\mu\nu}$ was
discussed in \cite{3}.  The relation was generalized to spinning spaces,
but only at the expense of introducing torsion.  The physical
interpretation of the dual metrics was not  clarified \cite{3}.
The geometrical interpretation of Killing tensors was investigated
in\cite{4}.

 Recently the structural equations for a Killing tensor of
  order two was investigated and the geometric duality between
  $g_{\mu\nu}$ and a non-degenerate  Killing tensor $K_{\mu\nu}$ was
analyzed in\cite{5}.

 An interesting example  of an Einstein's metric which admits
Killing-Yano tensors is Taub-NUT metric.  Taub-NUT metric is involved
in many mo\-dern studies in physics. For example the Kaluza-Klein
monopole of Gross and Perry \cite{6} and of Sorkin \cite{7} was
obtained by embedding the Taub-NUT gravitational instanton into
five-dimensional Kaluza-Klein theory.  Remarkably, the same object has
re-emerged in the study of monopole scattering.  In the long distance
 limit, neglecting radiation, the relative motion of the BPS monopoles
is described by the geodesics of this space \cite{8}\cite{9}. The
dynamics of well-separated monopoles is completely soluble and has a
Kepler type symmetry \cite{10,11,12,13}.

 The geodesic motion of pseudo-classical spinning
particles in Euclidian Taub-NUT were analyzed in
\cite{14} and the symmetries of extended Taub-NUT metrics recently
were studied in \cite{15,16,17}.
 Taub-NUT metric admits four Killing-Yano tensors which generate
four non-degenerate Killing tensors\cite{3}.
 On the other hand  for a given  manifold  $g_{\mu\nu}$ which
admits a non-degenerate Killing tensor  $K_{\mu\nu}$
two types of dual  metrics exist \cite{5}.
 An interesting question is to investigate the connection between
the symmetries of dual Taub-NUT metrics and the symmetries of
Taub-NUT metric.

 For these reasons the symmetries of the Taub-NUT dual metrics will be
analyzed.  The aim of this paper is to investigate
the {\it generic} and {\it non-generic} symmetries corresponding to
geodesic motion of pseudo-classical spinning particles on the  Taub-NUT
dual metrics.We will investigate the existence of the Runge-Lenz
symmetry for the dual metrics.

 The organization of the paper is as follows.
In Section 2 the geometric duality is presented .
  In Section 3 we investigate the symmetries
corresponding to Taub--NUT dual metrics and we construct the spinning
space. In Section 4 we present our conclusions.

 In Appendix 1  we write down Christoffel symbols and the scalar
curvature for two interesting  dual metrics. The calculus for all
Taub-NUT dual metrics were done, but due to their huge and
complicated expressions we cannot write them out in this paper.
 In Appendix 2 we present two plots of scalar curvature of two
interesting  dual metrics.
 %%%%%%%%%%%%%%%%

\section{Geometric Duality}

The equation of motion
of a particle on a geodesic is derived from the action
\begin{equation}
S=\int{d\tau \frac{1}{2} g_{\mu\nu}\dot{x^{\mu}}{\dot x^{\nu}}}
\end{equation}
The Hamiltonian has the form
$H=\frac{1}{2}g_{\mu\nu}p^{\mu}p^{\nu}$
where the Poisson brackets are
$\{x_{\mu}, p^{\nu}\}=\delta^{\nu}_{\mu}$.

  Let  us suppose that the metric $g_{\mu\nu}$ admits
a Killing tensor field $K_{\mu\nu}$.  A Killing tensor is a symmetric
tensor which satisfies the following relation:

\begin{equation}
D_{\lambda}K_{\mu\nu} +D_{\mu}K_{\nu\lambda} +D_{\nu}K_{\lambda\mu} = 0
\end{equation} where $D_{\mu}$ denote covariant derivatives.  {}From
  the covariant components $K_{\mu\nu}$ of the Killing tensor one can
construct a constant of motion $K=\frac{1}{2}K_{\mu\nu}p^{\mu}p^{\nu}$.
It can  be easy verified that $\{H,K\}=0$.

The formal similarity between the constants of motion H and K ,
and the symmetrical  nature of the condition implying the existence
of the Killing tensor amount to a reciprocal relation between two
different models:the model with Hamiltonian H and constant of motion K,
and a model with constant of motion H and Hamiltonian K.The relation
between the two models has a geometrical interpretation:
it implies that
if $K_{\mu\nu}$ are the contravariant components of a Killing tensor with
respect to the metric $g_{\mu\nu}$, then $g_{\mu\nu}$ must represent a
Killing tensor with respect to the metric defined by $K_{\mu\nu}$.
 When $K_{\mu\nu}$ has an inverse we interpret it as the
metric of another space and we can define the associated
Riemann-Christoffel connection $\hat\Gamma_{\mu\nu}^{\lambda}$ as usual
through the metric postulate
${\hat D}_{\lambda}K_{\mu\nu}=0$.
Here ${\hat D}$ represents the
covariant derivative with respect to $K_{\mu\nu}$.

This reciprocal relation between the metric structure of pairs of
spaces constitutes a duality relation: performing the operation of
 mapping a Killing tensor to a metric twice leads back to the original
theory.

 The geometric duality between $g_{\mu\nu}$ and a Killing
tensor $K_{\mu\nu}$ was analyzed in \cite{5}.In this case
 Killing's vectors equations in the dual space have the
following form \cite{5}
\begin{equation}\label{kiltex}
D_{\mu}\hat\chi_{\nu} + D_{\nu}\hat\chi_{\mu}
+2K^{\delta\sigma}(D_{\delta}K_{\mu\nu}){\hat\chi_{\sigma}}=0
\end{equation}
 Here $\hat\chi_{\sigma}$ are Killing vectors in dual spaces.

Let us suppose  that  metric  $g_{\mu\nu}$ admits a  Killing-Yano tensor
$f_{\mu\nu}$.
A Killing-Yano tensor is an antisymmetric tensor \cite{2}
which satisfies $D_{\mu}f_{\nu\lambda} + D_{\nu}f_{\mu\lambda}=0$

The corresponding Killing-Yano equations in the dual space has the
form\cite{5}
 \begin{equation}\label{anoua}
 D_{\mu}{\hat f_{\nu\lambda}} + D_{\nu}{\hat f_{\mu\lambda}} +{\hat f_{\nu}^{\delta}}D_{\delta}K_{\mu\lambda} +2
{\hat f_{\lambda}}^{\sigma}D_{\sigma}K_{\nu\mu} +
{\hat f_{\mu}}^{\delta}D_{\delta}K_{\nu\lambda} =0
\end{equation}
where ${\hat f_{\mu\nu}}$ is a Killing-Yano tensor on the dual
manifold. D represents the
covariant derivative with respect to $g_{\mu\nu}$.

%%%%%%%%%%%%%%%%%%%%%%%%%%%%%%

\section{Symmetries of the dual metrics for the self-dual Euclidean
Taub-NUT metric}

 The four-dimensional Taub-NUT metric
depends on a parameter $m$ which can be positive or negative, depending
on the application; for $m>0$ it represents a nonsingular solution of
the self-dual Euclidean equation and as such is interpreted as a
gravitational instanton. The standard form of the line element is

\begin{eqnarray}\label{ade}
ds^2 &=& \left(1+\frac{2m}{r}\right)(dr^2+r^2d\theta^2+r^2\sin^2\theta d\varphi^2)\nonumber\\ &&
+\frac{4m^2}{1+2m/r}(d\psi + \cos\theta d\varphi)^2
\end{eqnarray}

The Killing vectors for the metric (\ref{ade}) have the following form:
\begin{equation}
D^{(\alpha )}=R^{(\alpha )\mu }\,\partial _\mu ,~~~\alpha =1,\cdots ,4
\end{equation}
where
\begin{eqnarray}\label{ultima}
D^{(1)} &=&{\frac \partial {\partial \psi }}\label{vec14}   \\
D^{(2)} &=&-{\frac \partial {\partial \varphi }}\label{vec15} \\
D^{(3)} &=&\sin \varphi \,{\frac \partial {\partial \theta }}+\cos \varphi
\,\cot \theta \,{\frac \partial {\partial \varphi }}-{\frac{\cos \varphi }{%
\sin \theta }}\,{\frac \partial {\partial \psi }}   \\
D^{(4)} &=&-\cos \varphi \,{\frac \partial {\partial \theta }}+\sin \varphi
\,\cot \theta \,{\frac \partial {\partial \varphi }}-{\frac{\sin \varphi }{\sin \theta }}\,{\frac \partial {\partial \psi }}
\end{eqnarray}

The metric (\ref{ade}) admits four Killing-Yano tensors\cite{18}.
Three of these,
denoted by $f_i$ are special because they are covariant constant. In the
two-form notation the explicit expressions are
\begin{equation}\label{kyf}
f_i=4m(d\psi +\cos \theta d\varphi )dx_i-\epsilon _{ijk}(1+{\frac{2m}r}%
)dx_j\wedge dx_k
\end{equation}
where the $dx_i$ are standard expressions in terms of the 3-dimensional
spherical co-ordinates $(r,\theta ,\varphi )$.
The fourth Killing-Yano tensor has the following form
\begin{equation}\label{ky}
Y=4m(d\psi
+\cos \theta d\varphi )\wedge dr+4r(r+m)(1+{\frac r{2m}})\sin \theta
d\theta \wedge d\varphi
\end{equation}

 Now we would like to investigate the symmetries of the dual
Taub-NUT metrics.

We will construct firstly the Taub-NUT dual metrics using
 the geometric
duality between $g_{\mu\nu}$ and  $K_{\mu\nu}$ \cite{5}.
 We know
 that  when a manifold M admits a Killing-Yano tensor $f_{\mu\nu}$  we
can construct a corresponding  Killing tensor\cite{14} $K_{\mu\nu}$
 as
 \begin{equation}\label{eky}
K_{\mu\nu}=f_{\mu}^{\lambda}f_{\nu\lambda}
\end{equation}

  Using (\ref{eky}) and
(\ref{ky}) the line element for the  dual metric becomes

\begin{eqnarray}\label{kilten}
dk^{2} &=&
\left(1+\frac{2m}{r}\right)\left(dr^2+\frac{r^2}{m^2}(r+m)^2
(d\theta^2+\sin^2\theta d\varphi^2)\right) \nonumber\\ &&
+\frac{4m^2}{1+2m/r}(d\psi + \cos\theta d\varphi)^2.
\end{eqnarray}

 The Taub-NUT metric admits three more second-rank Killing tensors
 of the form \cite{18}
 \begin{equation}\label{ru}
 K_{\mu\nu}^{(i)}=f_{\mu\lambda}^{(i)}Y^{\lambda}_{\nu}+
 f_{\nu\lambda}^{(i)}Y^{\lambda}_{\mu}
 \end{equation}
 Here $f_{\mu\nu}^{i}$ and $Y_{\mu\nu}$ are given by
 (\ref{kyf}) and(\ref{ky}).
  They form a conserved vector of Runge-Lenz type.
 Using (\ref{ru}) the line element for the dual metrics
becomes
 \begin{eqnarray}\label{taubi}
dk^{2}_{(i)} &=&
-\frac{2}{m}\left(1+\frac{2m}{r}\right)\left(1+\frac{m}{r}\right)
r_i(dr^2+r^2d\theta^2+r^2\sin^2\theta d\varphi^2)+\nonumber\\ &&
\frac{8m^2}{r(1+2m/r)}r_i(d\psi+\cos\theta d\varphi)^2
+\frac{2r}{m}\left(1+\frac{2m}{r}\right)^2drdr_i \nonumber\\ &&
+4\left(1+\frac{2m}{r}\right)({\bf r}\times d{\bf r})_i(d\psi
+\cos\theta d\varphi)  \end{eqnarray}

 We found  that the dual metric (\ref{kilten})
    has the same Killing vectors like Taub-NUT metric (\ref{ade})
because relation (\ref{kiltex}) is identically satisfied.

For $i=3$  the corresponding metric from  (\ref{taubi})
admits two Killing vectors (\ref{vec14}) and (\ref{vec15})
but  for $i=1,2$  we found from (\ref{kiltex})
only one Killing vector (\ref{vec14}).

Because the  Weyl tensor of (\ref{kilten}) has non-vanishing components
\\ (e.g $C_{1234}=2\,({m}^{2}\,{r}\,{\rm sin}(\,{ \theta}\,)\,(\,2\,{r}
+ 3\,{m}\,)/(\,{r} + 2\,{m}\,)^{2}\,(\,{r} + {m}\,)$)
 , the metric is not conformally flat.  We have obtained
 the same result for (\ref{taubi}) but the expressions are too long to
 be provided here.

Now we would like to investigate the Killing-Yano tensors of order two
for the dual metrics (\ref{kilten},\ref{taubi}).

We have six
independent components of Killing-Yano tensor $f_{\mu\nu}$ and 24
independent equations (\ref{anoua}).

 Replacing $D_{\mu}f_{\lambda\gamma}={\partial_{\mu}
f_{\lambda\gamma}}- f_{\delta\gamma}\Gamma_{\lambda\mu}^{\delta} -
f_{\lambda\delta}\Gamma_{\mu\gamma}^{\delta}$  in (\ref{anoua})
and using the corresponding  expressions of Christoffel's symbols for
(\ref{kilten})  we get a set of Killing-Yano
equations.
We found that
(\ref{anoua}) has no solution, therefore
the dual metrics (\ref{kilten}) and (\ref{taubi}) have no
extra symmetry of Runge-Lenz type.

 In \cite{3} using
geometric duality between $g^{\mu\nu}$ and $K^{\mu\nu}$ four
dual Tab-NUT metrics were found.
 The inverse matrix of the covariant form from (\ref{kilten}) give us
the dual line element
 \begin{eqnarray}\label{elem}
 d\tilde{s}^2 &=&
\left(1+\frac{2m}{r}\right)\left(dr^2+\frac{m^2r^2}{(r+m)^2}
(d\theta^2+\sin^2\theta d\varphi^2)\right)\nonumber\\ &&
+\frac{4m^2}{1+2m/r}(d\psi + \cos\theta d\varphi)^2.
\end{eqnarray}
 and the dual metrics corresponding to Runge-Lenz vector have the
 following form  (for more details see \cite{3}).

 \begin{eqnarray}\label{31eq} d\tilde{s}^2_{(i)}
 &=& \frac{-1}{r^2_i-(r+2m)^2}\{-\frac{2m^2}{r}\left(1+
 \frac{2m}{r}\right )r_i (dr^2+r^2d\theta^2+r^2\sin^2\theta d\varphi^2)
\nonumber\\ && +\frac{8m^3(1+m/r)}{(1+2m/r)}r_i(d\psi+\cos\theta
d\varphi)^2 +2mr\left(1+\frac{2m}{r}\right )^2drdr_i \nonumber\\ &&
+4m^2\left(1+\frac{2m}{r}\right)({\bf r}\times d{\bf r})_i(d\psi
+\cos\theta d\varphi)\}
\end{eqnarray}
 Next step is to investigate the symmetries of the
metrics (\ref{elem},\ref{31eq}).

If we make the transformation to a new variable u as
$u=re^{\frac{r}{m}}$  the line element (\ref{elem}) becomes

\begin{equation}\label{alta}
 d\tilde{s}^{2}=F(u)(du^{2} +u^{2}(d\theta^{2}
+\sin^{2}\theta d\varphi^{2})) +G(u)(d\psi +\cos\theta d\varphi^{2})
\end{equation}
where $F(u)$ and $G(u)$ are given by
$F(re^{\frac{r}{m}})=\frac{e^{-\frac{2r}{m}}}{\left( \frac{1}{m}
+\frac{1}{r}\right)^{2}r^{2}}\left( 1+\frac{2m}{r}\right),\\
G(re^{\frac{r}{m}})=\frac{4m^2}{1+\frac{2m}{r}}$.
The metric (\ref{alta})
is a particular form of extended Taub-NUT metric presented in
\cite{15,16}.
On the other hand, in the case of geometric duality between $g^{\mu\nu}$
 and $K^{\mu\nu}$ the  equations (\ref{kiltex}) and (\ref{anoua}) are
not valid (for more details see \cite{5}). Because of this the
 symmetries of the metrics (\ref{elem},\ref{31eq}) will be investigated
 using (see \cite{2,3})
 the Killing vectors equations \begin{equation}\label{kilnou}
 {\hat D_{\mu}}{\hat\chi_{\nu}} +{\hat D_{\nu}}{\hat\chi_{\mu}}=0
\end{equation}

and  the Killing-Yano equations
\begin{equation}\label{yanotex}
\{\mu\nu\lambda\}={\hat D_{\mu}}{\hat f_{\nu\lambda}} +
{\hat D_{\nu}}{\hat f_{\mu\lambda}}=0 \end{equation}
 Solving (\ref{kilnou}) we found
that the dual metric (\ref{elem}) admits the same Killing vectors as
metric (\ref{ade}).This result is in agreement with those
from\cite{15}.
 For $i=3$, (\ref{kilnou}) give us for the corresponding metric
in (\ref{31eq}) two Killing vectors (\ref{vec14}) and
(\ref{vec15}). In the case $i=1,2$ from (\ref{31eq}) we found
  only one Killing vector
(\ref{vec14}) for  the corresponding metrics.

 Now we investigate if the dual metrics (\ref{elem},\ref{31eq})
admits Killing-Yano tensors of order two.
Our  strategy is quite straightforward in this case.
We simply write down all the components of the equation (\ref{yanotex})
explicitly.
The total number of components of $f_{\mu\nu}$ is six,
 while the number of independent equations for  eq.(\ref{yanotex}) is
 24.

\begin{eqnarray}\label{tabel}
 &\{rr \theta\}=0&,\{rr\varphi\}=0,\{rr
\psi\}=0,\{\theta\theta r\}=0,\{\theta\theta\varphi\}=0\cr
 &\{\theta\theta\psi\}=0&
,\{\varphi\varphi r\}=0,\{\psi\psi r\}=0,\{\psi\psi\theta\}=0,
\{\psi\psi\varphi\}=0\cr
&\{r\theta\varphi\}=0&,\{r\theta\psi\}=0,\{r\varphi\theta\}=0,\{r\varphi\psi\}=0,\{r
\psi\theta\}=0\cr
 &\{r\psi\varphi\}=0&,\{\theta\varphi
r\}=0,\{\theta\psi
r\}=0,\{\varphi\theta\psi\}=0,\{\varphi\varphi\theta\}=0\cr
&\{\psi\theta\varphi\}=0&,\{\psi\varphi r\}=0,\{\psi\varphi\theta\}=0,\{\varphi\varphi\psi\}=0
\end{eqnarray}

Solving (\ref{tabel}) we found no solution for the metrics
 (\ref{elem}) and (\ref{31eq}). Then the dual metrics
 (\ref{elem},\ref{31eq}) have no Killing-Yano tensors and Runge-Lenz
 type symmetry.
 The Weyl tensor has non-vanishing components for
 (\ref{elem})(e.g. $ C_{1234}=-2\,{\displaystyle {m}^{3}\,{r}\,{\rm
sin}(\,{ \theta}\, )/(\,{r} + {m}\,)\,(\,{r} + 2\,{m}\,)^{2}}$) and then
 the metric in not conformally flat.
 Metrics (\ref{31eq})
are not conformally flat because they have non-vanishing Weyl tensor
components,
but all expressions are too long to be written here.

 %%%%%%%%%%%%%%%%%
\subsection{Generic and non-generic symmetries}

 An action for the geodesic of spinning space is:
\begin{equation}\label{spin}
S=\int_a^bd\tau \left( \,{\frac 12}\,g_{\mu \nu }(x)\,\dot{x}^\mu \,\dot{x}%
^\nu \,+\,{\frac i2}\,g_{\mu \nu }(x)\,\psi ^\mu \,{\frac{D\psi ^\nu }{D\tau
}}\right) .
\end{equation}
 The overdot denotes an ordinary proper-time
derivative $d/d\tau $, whilst the covariant derivative of a Grassmann
variable $\psi ^\mu $ is defined by ${\frac{D\psi ^\mu
}{D\tau }}=\dot{\psi}^\mu +\dot{x}^\lambda \,\Gamma _{\lambda \nu }^\mu
\,\psi ^\nu$ .
 In general, the symmetries of a spinning-particle model can
be divided into two classes.
 First, there are conserved quantities
which exist in any theory and these are called {\it  generic}
constants of motion .
 It was shown that for a spinning particle model defined by
the action (\ref{spin}) there are four generic symmetries \cite{14}.

\begin{enumerate}
\item  {Proper -time translations and the corresponding constant of motion
are given by the Hamiltonian:
\begin{equation}\label{prime}
H={\frac 12}\,g^{\mu \nu }\,\Pi _\mu \,\Pi _\nu
\end{equation}
}

\item  {Supersymmetry generated by the supercharge
\begin{equation}
Q=\Pi _\mu \,\psi ^\mu
\end{equation}
}

\item  { Chiral symmetry generated by the chiral charge
\begin{equation}
\Gamma _{*}={\frac 1{4!}}\sqrt{-g}\epsilon _{\mu \nu \lambda \sigma }\,\psi
^\mu \,\psi ^\nu \,\psi ^\lambda \,\psi ^\sigma
\end{equation}
}

\item  {Dual supersymmetry, generated by the dual supercharge
\begin{equation}\label{ultim}
Q^{*}={\frac 1{3!}}\,\sqrt{-g}\epsilon _{\mu \nu \lambda \sigma }\,\Pi ^\mu
\,\psi ^\nu \,\psi ^\lambda \,\psi ^\sigma .
\end{equation}
}
Here $\Pi _\mu =g_{\mu \nu }\dot{x}^\mu $ represents the covariant momentum.
\end{enumerate}

The second kind of conserved quantities, called
{\it non-generic}, depend on the explicit form of the metric $g_{\mu
\nu }(x)$.
    {\it Non-generic} symmetries are associated with the existence of
Killing- Yano tensors on a given manifold.
 The existence of a Killing-Yano tensor $f_{\mu\nu}$ of the bosonic
manifold is equivalent to the existence of a supersymmetry for the
spinning particle with supercharge

$Q_f=f^{\mu}_{a}\Pi_{\mu}\psi^{a}-\frac{1}{3}iH_{abc}\psi^{a}\psi^{b}\psi^{c}$
satisfies
$\{Q,Q_f\}=0$,
where $H_{\mu\nu\lambda}= D_{\lambda}f_{\mu\nu}$.

 For the metric (\ref{kilten}) we found  the {\it generic} symmetries
(\ref{prime}---\ref{ultim}) in the following form

\begin{eqnarray}
 \lefteqn{H=  -\,{\displaystyle \frac {1}{2}}\,
{\displaystyle \frac {(\, -{r}^{2}\,{m}^{2} - 4\,{r}\,{m}^{3}
 - 4\,{m}^{4}\,)\,{\dot{r}}^{2}}{{m}^{2}\,(\,{r} + 2\,{m}\,)\,{r} }}}
\nonumber\\ & &
-{\displaystyle \frac {1}{2}}\,{\displaystyle \frac
{(\, -6\,{r}^{5}\,{m} - 12\,{r}^{3}\,{m}^{3} - 13\,{r}^{4} \,{m}^{2} -
{r}^{6} - 4\,{m}^{4}\,{r}^{2}\,)\,{ \dot{\theta}}^{2}}{{ m}^{2}\,(\,{r}
+ 2\,{m}\,)\,{r}}} \nonumber \\ & &
{\displaystyle -\frac {1}{2}} \left( {\vrule
 height0.44em width0em depth0.44em} \right. \! \!
- 13\,{r}^{4}\,{m}^{2}\,{\rm sin}(\,{ \theta}\,)^{2}
- 6\,{r}^{5}\,{m}\,{\rm sin}(\,{ \theta}\,)^{2 } -
{r}^{6}\,{\rm sin}(\,{ \theta}\,)^{2}-4\,{m}^{4}\,{r}^{2}\nonumber \\ & &
 \mbox{} - 12\,{r}^{3}\,{m}^{3}\,{\rm sin}(\,{ \theta}\,)^{2} \!
\! \left.  {\vrule height0.44em width0em depth0.44em} \right) {
\dot{\phi}}^{ 2} \left/ {\vrule height0.37em width0em depth0.37em}
\right. \!  \!  (\,{m}^{2}\,(\,{r} + 2\,{m}\,)\,{r}\,)\nonumber \\ & &
\mbox{} +
 4\,{\displaystyle \frac {{m}^{2}\,{\rm cos}(\,{ \theta}\,)\,{r}\,{
\dot{\phi}}\,{ \dot{\psi}}}{{r} + 2\,{m}}} + 2\, {\displaystyle \frac
{{r}\,{m}^{2}\,{ \dot{\psi}}^{2}}{{r} + 2\,{m }}}
\end{eqnarray}

\begin{eqnarray}
\lefteqn{Q =  \left( \! \,1 + 2\,{\displaystyle \frac {{
m}}{{r}}}\, \!  \right) \,{\dot{r}}\,{ \psi^{r}} + {\displaystyle
\frac {{r}^{2}\, \left( \! \,1 + 2\,{\displaystyle \frac {{m}}{{r
}}}\, \!  \right) \,(\,{r} + {m}\,)^{2}\,{\dot{\theta}}\,{
\psi^{\theta}}}{{m} ^{2}}} + } \nonumber \\ & &
 \left( \! \, \left( \!
 \,{\displaystyle \frac {{r}^{2}\,(\, {r} + {m}\,)^{2}\, \left( \! \,1
+ 2\,{\displaystyle \frac {{m}}{ {r}}}\, \!  \right) \,{\rm sin}(\,{
\theta}\,)^{2}}{{m}^{2}}} + 4 \,{\displaystyle \frac {{m}^{2}\,{\rm
cos}(\,{ \theta}\,)^{2}}{1 + 2\,{\displaystyle \frac {{m}}{{r}}}}}\, \!
 \right) \,{\dot{\phi}} + 4\,{\displaystyle \frac {{m}^{2}\,{\rm
 cos}(\,{ \theta}\,)\, {\dot{\psi}}}{1 + 2\,{\displaystyle \frac
{{m}}{{r}}}}}\, \!  \right) {\psi^{\varphi}} \nonumber \\ & &
 \mbox{} +  \left(
 \! \,4\,{\displaystyle \frac {{m} ^{2}\,{\rm cos}(\,{
\theta}\,)\,{\dot{\phi}}}{1 + 2\,{\displaystyle \frac {{m}}{{r}}}}} +
4\,{\displaystyle \frac {{m}^{2}\,{\dot{\psi}} }{1 + 2\,{\displaystyle
\frac {{m}}{{r}}}}}\, \!  \right) \,{ \psi^{\psi}}
\end{eqnarray}

\begin{equation}
\Gamma_{*}=\frac{1}{12m}(r + 2m)r(r+m)^{2}\sin(\theta)^2
 \psi ^r\,\psi ^\theta \,\psi ^{\varphi}\,\psi ^\psi
\end{equation}

\begin{eqnarray}
Q^{*}&=&\frac{(r+2m)r(r+m)^{2}\sin(\theta)^2}{3m}(\dot{r}\,\psi ^\theta 
\,\psi ^{\varphi}\,\psi ^\psi -\dot{\theta}\,\psi ^r\,\,\psi 
^{\varphi}\,\psi ^\psi \nonumber \\ & &
 + \dot{\varphi}\,\psi 
^r\,\psi^\theta \,\psi ^\psi \, -\dot{\psi}\,{\psi ^r} \,\psi ^\theta 
 \,\psi^{\varphi})
 \end{eqnarray}

Because the determinants of all dual
metrics (\ref{taubi},\ref{elem},\ref{31eq}), do not vanish we have
similar expressions, in corresponding cases, for the {\it generic}
symmetries\\ (\ref{prime}---\ref{ultim}).

  As it was shown above the  dual metrics  have no
 Killing-Yano tensors, and because of this we have no
{\it non-generic} symmetries for metrics (\ref{kilten},
\ref{taubi},\ref{elem},\ref{31eq}).

%%%%%%%%%%%%%%%%%%%%%%%%%%%%%%%%%%%%%%%%%%%%%%%%%%%%%%%%%%%%%%%

\section{Conclusions}

  Recently geometric duality was analyzed for a metric which admits
a non-degenerate Killing tensor of order two \cite{3,5}.
 For a given manifold which admits a non-degenerate Killing tensor of
order two, geometric duality give us two types of dual metrics \cite{5}.
 An interesting example arises
when the manifold admits Killing-Yano tensors because they generate
Killing tensors.

 In this paper the symmetries of the dual Taub-NUT metrics were
investigated.
  Taub-NUT metric admits four  Killing-Yano tensors of order
two and we have four corresponding dual metrics.
 We have obtained that the number of Killing vectors of the dual
 Taub-NUT metrics depend drastically on their particular form.
 We found that metrics (\ref{kilten},\ref{elem}) have the same Killing
vectors as the Taub-NUT metric (\ref{ade}), the corresponding dual
metrics for $i=3$ in (\ref{taubi},\ref{31eq}) have two Killing vectors
 (\ref{vec14},\ref{vec15}),
and for $i=1,2$ we get only one Killing vector (\ref{vec14}) .
 We have obtained, by solving Killing-Yano equations, that  all
dual Taub-NUT metrics do not have Killing-Yano tensors.
 This means that all dual metrics do no admit extra
symmetries of Runge-Lenz type.

The scalar curvature of Taub--NUT metric (\ref{ade}) is zero,
but the corresponding dual metrics
 (\ref{kilten},\ref{taubi},\ref{elem},\ref{31eq}) have non
vanishing scalar curvatures and no Runge-Lenz vector.
  We found that the scalar curvature of dual metric (\ref{kilten}) is
positive and the corresponding scalar curvature for (\ref{elem}) is
negative.

  We have obtained that the dual metric (\ref{elem}) is a
special case of extended Taub-NUT metric given in \cite{15}. We
obtained also that all dual Taub-NUT metrics  are not conformally
flat.
 The spinning space was constructed and the {\it generic}
and {\it non-generic} symmetries of the dual Taub-NUT metrics were
analyzed.
 We found that dual Taub-NUT metrics have not {\it non-generic}
symmetries. Our result differs from those presented in \cite{3}.
 Geometric duality, in case of Taub-NUT metric, reduces the number of
  symmetries of the dual metrics.
  Finding the number of
 symmetries of dual metrics for a given manifold which admits
 a Killing tensor is an interesting problem,
and it requires further investigation.

%%%%%%%%%%%%%%%%%%%%%%%%%%%%%%%%%%%%%%%%%%%%%%%
\vskip 1cm
\newpage

{\large\bf Acknowledgments}:
\bigskip

 One of the authors (D.B.) thanks Prof.S.Ba\. za\' nski and Prof. S.
Manoff, for
helpful discussions and for continuous encouragements.

%\vfill\eject

\newpage

\begin{center}
{\bf\large Appendix 1.}
\end{center}
\vskip .5cm

For the metric (\ref{kilten})  non-vanishing Christoffel coefficients  are\\

$$\Gamma_{11}^{1}=-\frac{m}{r(r+2m)},~~~\Gamma_{12}^{2}=-\frac{m^2+4mr+2r^2}{r(r^2+3rm+2m^2)}$$
$$\Gamma_{13}^{3}=\frac{m^2+4rm+2r^2}{r(r^2+3rm+2m^2)},~~\Gamma_{13}^{4}=-\frac{\cos(\theta)(3m+2r)}{r^2+3rm+2m^2}$$
$$\Gamma_{14}^{4}=\frac{m}{r(r+2m)},~~
\Gamma_{22}^{1}=-\frac{r(r+m)(m^2+4rm+2r^2)}{(r+2m)m^2}$$
$$\Gamma_{23}^{3}=\frac{\cos(\theta)(r^4+6r^3m+13r^2m^2+12rm^3+2m^4)}{\sin(\theta)(r^4+6r^3m+13r^2m^2+12rm^3+4m^4)}$$
$$\Gamma_{23}^{4}=-\frac{(\cos^2(\theta)+1)(r^4+6mr^3+13r^2m^2+12rm^3)+4m^4}
{2\sin(\theta)(r^4+6r^3m+13r^2m^2+12rm^3+4m^4)}$$
$$\Gamma_{24}^{3}=-2\frac{m^4}{\sin(\theta)(r^4+6r^3m+13r^2m^2+12rm^3+4m^4)}$$
$$\Gamma_{24}^{4}=2\frac{m^4\cos(\theta)}{\sin(\theta)(r^4+6r^3m+13r^2m^2+12rm^3+4m^4)}$$
$$\Gamma_{33}^{1}=-\frac{r(2r^5\sin^2(\theta)+14r^4m\sin^2(\theta)+
37r^3m^2\sin^2(\theta)+45r^2m^3\sin^2(\theta)+24rm^4\sin^2(\theta)+4m^5)}
{m^2(r^3+6r^2m+12rm^2+8m^3)}$$
$$\Gamma_{33}^{2}=-\frac{r\sin(\theta)\cos(\theta)(r^3+6r^2m+13rm^2+12m^3)}{\sin(\theta)(r^4+6r^3m+13r^2m^2+12rm^3+4m^4)}$$
$$\Gamma_{34}^{1}=-4\frac{m^3r\cos(\theta)}{r^3+6r^2m+12rm^2+8m^3}$$
$$\Gamma_{34}^{2}=2\frac{m^4\sin(\theta)}{r^4+6r^3m+13r^2m^2+12rm^3+4m^4},~~\Gamma_{44}^{1}=-4\frac{m^3r}{(r+2m)^3}$$
\vskip .5cm
and the curvature\\

$$R=2\frac{6m^3+21rm^2+22r^2m+8r^3}{2m^5+9rm^4+16r^2m^3+14m^2r^3+6r^4m+r^5}$$
\vskip .5cm

For the metric (\ref{elem})
%$$d\tilde{s}^2=\left(1+\frac{2m}{r}\right)\left(
%dr^2+\frac{m^2r^2}{(r+m)^2}(d\theta^2+\sin^2\theta
%d\varphi^2)\right)+\\
%\frac{4m^2}{1+2m/r}(d\psi+\cos\theta d\varphi)^2.$$
non-vanishing Christoffel components are\\
%\vskip .5cm
$$\Gamma_{11}^{1}=-\frac{m}{r(r+2m)},~~\Gamma_{12}^{2}=\frac{m^2}{r(r+m)(r+2m)}$$
$$\Gamma_{13}^{3}=\frac{m^2}{r(r^2+3rm+2m^2)},~~\Gamma_{13}^{4}=\frac{m\cos(\theta)}{r^2+3rm+2m^2},
~~\Gamma_{14}^{4}=\frac{m}{r(r+2m)}$$
$$\Gamma_{22}^{1}=-\frac{rm^4}{(r+m)^3(r+2m)},~~\Gamma_{23}^{3}=-\frac{(r^2-2m^2)\cos(\theta)}{\sin(\theta)(r+2m)^2}$$
$$\Gamma_{23}^{4}=\frac{3\cos^2(\theta)r^2-r^2-4\sin^2(\theta)rm-4m^2}{2\sin(\theta)(r+2m)^2}$$
$$\Gamma_{24}^{3}=-2\frac{(r^2+2rm+m^2)}{\sin(\theta)(r+2m)^2},~~
\Gamma_{24}^{4}=2\frac{\cos(\theta)(r^2+2rm+m^2)}{\sin(\theta)(r^2+4rm+4m^2)}$$
$$\Gamma_{33}^{1}=-\frac{rm^3(r^2m+11r^2m\cos^2(\theta)+4rm^2+8rm^2\cos^2(\theta)+4m^3+4r^3\cos^2(\theta))}
{r^6+9r^5m+33r^4m^2+63r^3m^3+66r^2m^4+36rm^5+8m^6)}$$
$$\Gamma_{33}^{2}=\frac{r(3r+4m)\cos(\theta)\sin(\theta)}{(r+2m)^2},~~
\Gamma_{34}^{1}=-4\frac{rm^3\cos(\theta)}{r^3+6r^2m+12rm^2+8m^3}$$
$$\Gamma_{34}^{2}=2\frac{(r+m)^2\sin(\theta)}{(r+2m)^2},~~\Gamma_{44}^{1}=-4\frac{m^3r}{(r+2m)^3}$$

\vskip .5cm
and the curvature is:
$$R=-2\frac{6m^2+3rm+2r^2}{m(r^3+4r^2m+5rm^2+2m^3)}$$
\vskip .5cm

\newpage
 
\begin{center}
{\bf\large Appendix 2.}
\end{center}

    \begin{figure}[h]
\centerline{\psfig{file=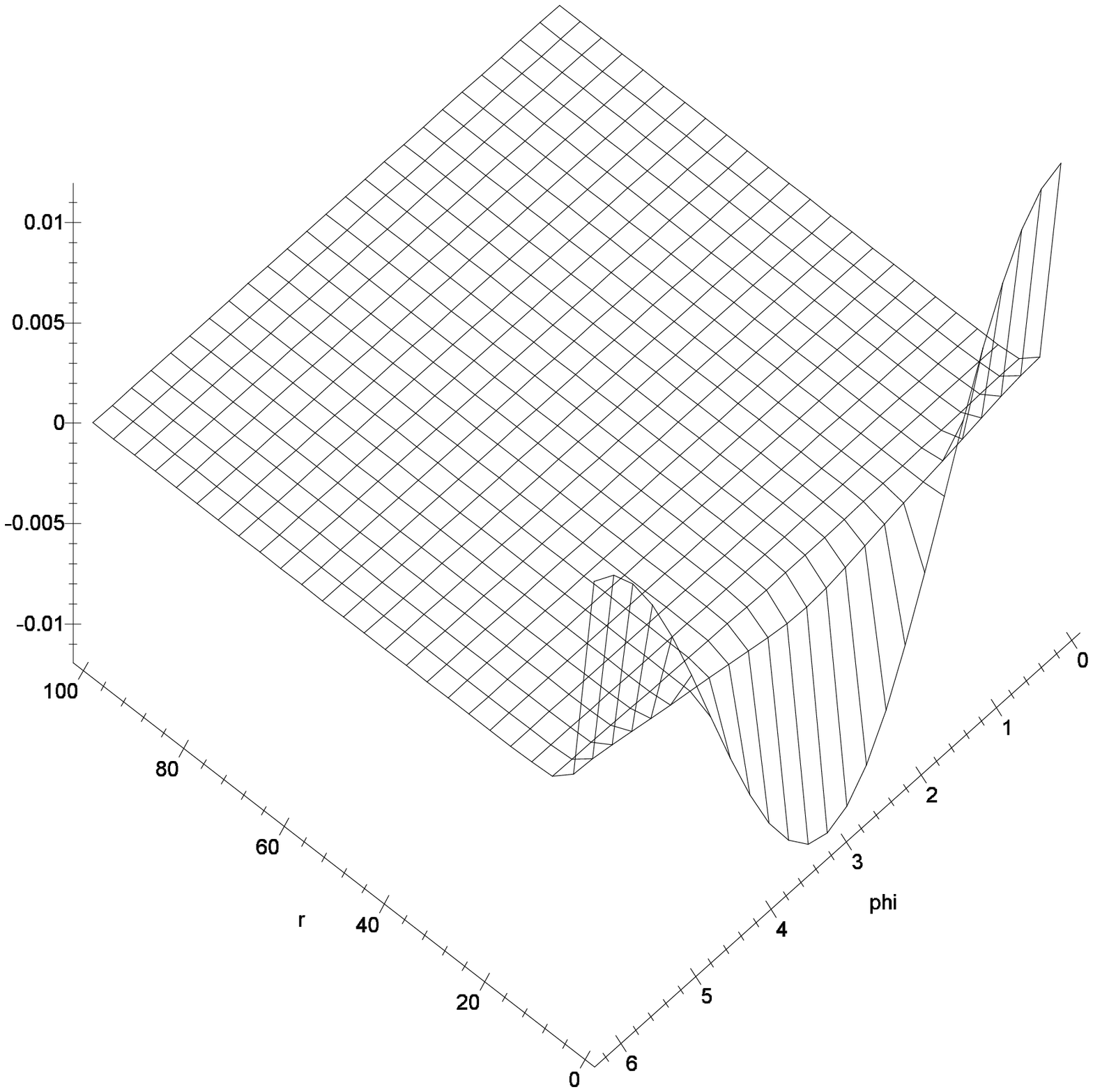, height=16cm}}
\caption{Ricciscalar corresponding to (\ref{taubi}) for i=1}
    \end{figure}

%\newpage
\begin{figure}[h]
\centerline{\psfig{file=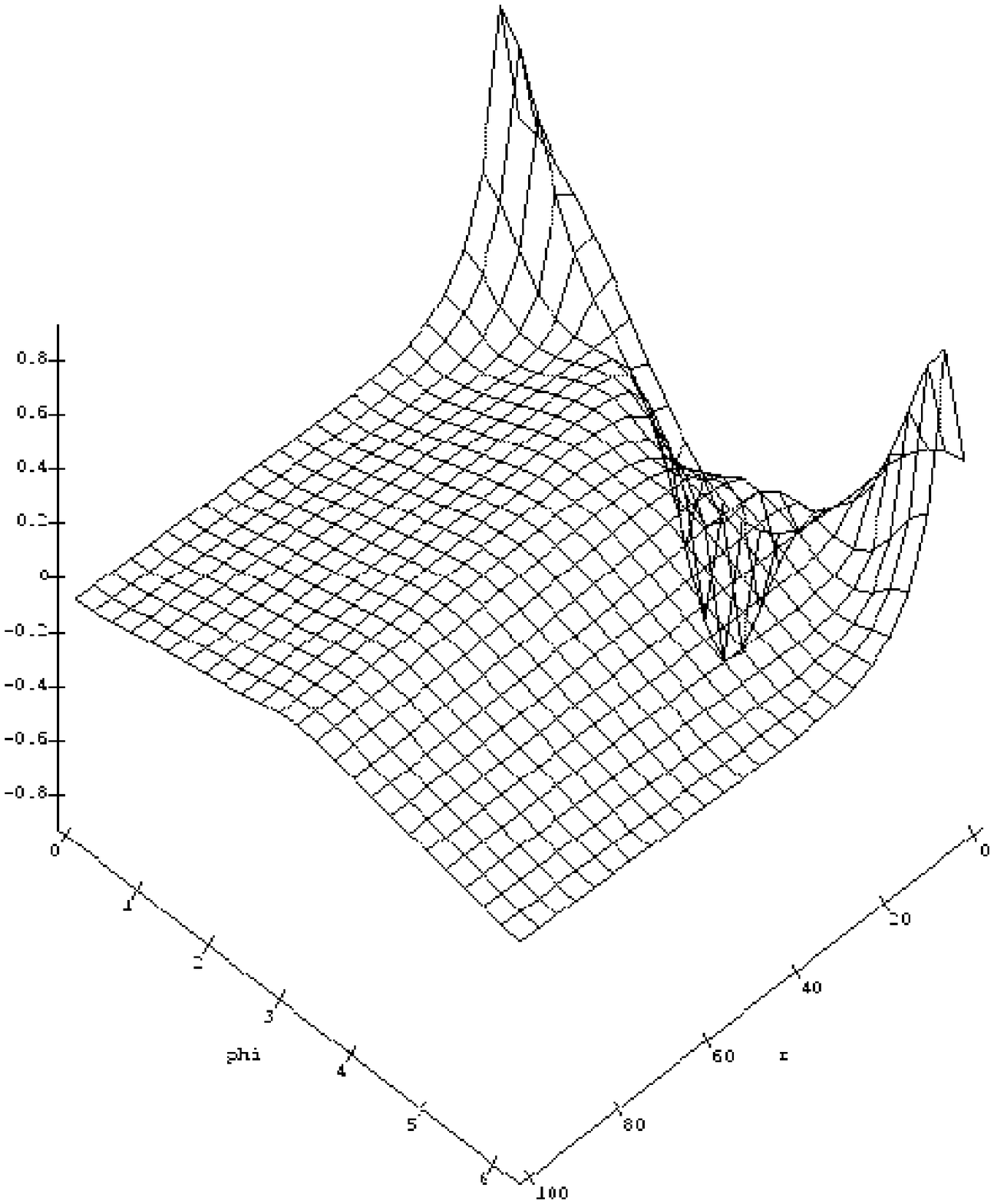, height=16cm}}
\caption{Ricciscalar corresponding to (\ref{31eq}) for i=1}
\end{figure}

%%%%%%%%%%%%%%%%%%%%%%%%%%%%%%%%%%%%%%%%%
%%%%%%%%%%%%%%%%%%%%%%%%%%%%%%%%%%%%%%%%%%%%%%%%%%%%%%%%%%%%%%%%
\end{document}